 \documentclass[preprint, prb, showpacs, superscriptaddress]{revtex4-1}

\usepackage{graphicx, epstopdf}
\usepackage{subfigure}
\usepackage{float}

\usepackage{amsmath}
\usepackage{amsfonts}
\usepackage{amssymb}
\usepackage{dcolumn}
\usepackage{bm}
\usepackage{subfigure}
\usepackage{caption}

\begin{document}

\title{Quantum Spin Hall Effect in 2D Transition Metal Dichalcogenide Haeckelites}

\author{S. M. Nie}
\author{Zhida Song}
\affiliation{Beijing National Laboratory for Condensed Matter Physics,
  and Institute of Physics, Chinese Academy of Sciences, Beijing
  100190, China}

\author{Hongming Weng}
 \email{hmweng@iphy.ac.cn}
 \affiliation{Beijing National Laboratory for Condensed Matter Physics,
  and Institute of Physics, Chinese Academy of Sciences, Beijing
  100190, China}
\affiliation{Collaborative Innovation Center of Quantum Matter,
  Beijing, 100084, China}
\author{Zhong Fang}
 \email{zfang@iphy.ac.cn}
 \affiliation{Beijing National Laboratory for Condensed Matter Physics,
  and Institute of Physics, Chinese Academy of Sciences, Beijing
  100190, China}
\affiliation{Collaborative Innovation Center of Quantum Matter,
  Beijing, 100084, China}

\date{\today}

\begin{abstract}
By using first-principles calculation, we have found that a family of 2D transition metal dichalcogenide haeckelites with square-octagonal lattice $MX_2$-4-8 ($M$=Mo, W and $X$=S, Se and Te) can host quantum spin hall effect. The phonon spectra indicate that they are dynamically stable and the largest band gap is predicted to be around 54 meV, higher than room temperature. These will pave the way to potential applications of topological insulators. We have also established a simple tight-binding model on a square-like lattice to achieve topological nontrivial quantum states, which extends the study from honeycomb lattice to square-like lattice and broads the potential topological material system greatly.
\end{abstract}

\maketitle

\section{introduction}

Inspired by the impressed progress in theory and applications, numerous
researchers turn their attentions to two-dimensional (2D) systems especially
miraculous graphene\cite{novoselov2004}. Recently, another 2D system, transition
metal dichalcogenides (TMDs) $MX_2$ with $M$=Mo, W, Ti, etc. and $X$=S, Se, Te,
has been widely explored due to wide range of electronic properties
~\cite{radisavljevic2013mobility,wang2012electronics, xu2014topological,shirodkar2014emergence,
cao2012valley} and easy fabrication. Bulk TMDs are composed of 2D X-M-X layers
stacked on top of each other. The bonding within those trilayer sheets is covalent
while the coupling between adjacent sheets is weak van der Waals (vdW) interaction.
2D TMD layers can be manufactured not only by mechanical\cite{novoselov2005,radisavljevic2011}
and chemical exfoliation\cite{eda2011,coleman2011} of their layered bulk counterparts,
but also by chemical vapor deposition (CVD)\cite{lee2012} or two-step thermolysis\cite{liu2012}.
However, defect will be inevitably produced during the manufacture, and it
will modify the electronic structure significantly in low dimensional system.

Graphene with 5-7 defects, which is often called Haeckelites in honor of
the German biologist and naturalist Ernst Haeckel, has been theoretically
proposed about twenty years ago\cite{crespi1996,terrones2000}. In contrast to
comprehensive understanding of the defect in graphene\cite{sheng2012octagraphene},
defect in 2D TMDs may just launch on. There are still so many issues need
to be tackled. W. Li {\it et al.}\cite{PhysRevB.89.205402} and H. Terrones {\it et al.}\cite{terrones2014} 
proposed that a new planar sheet could be generated from original hexagonal TMDs when introducing 4-8 defects. 
This new planar sheet is TMD Haeckelites with square-octagonal lattice.
The periodic 4-8 defects have been observed in the grain boundaries of MoS$_2$ and most probably will
exist in other TMDs\cite{van2013}, too. High-resolution transmission electron microscopy (HR-TEM)
is a powerful instrument to selectively suppress or enhance bond rotations
and produce defects in sample due to the ballistic procedure between high
energetic electrons and sample atoms. Using aberration corrected HR-TEM device,
a disordered graphene Haeckelite has been produced in situ\cite{kotakoski2011}.
Combing first-principles calculations and HR-TEM experiments, H. Komsa demonstrated
that it is possible to observe defect formation under exposure to an 80 keV electron
beam in MoS$_2$ system\cite{komsa2012}.


Recently, Qian et al.\cite{qian2014quantum} predicted that some 2D TMDs with 1$\text{T}'$ structure
can be large-gap 2D topological insulators (TIs) though most of them are in 1$\text{H}$ structure and are not TIs. 
It is natural to ask whether TI state can exist in the defected 2D TMDs based on 1$\text{H}$ structure.
2D TI \cite{kane2005,bernevig2006}, also known as quantum spin hall (QSH) insulator, was firstly
proposed in graphene, where spin-orbit coupling (SOC) opens a band gap at the Dirac point. 
On account of weak  SOC strength, the band gap is so small (order of $10^{-3}$ meV)\cite{yao2007spin}
that this proposal is hardly to be verified by experiments. So far, QSH effect
has only been observed in HgTe/CdTe\cite{konig2007} and InAs/GaSb\cite{knez2011} quantum wells. Both
of them require precisely controlled MBE growth and ultralow temperature. The study of 2D TI has been seriously 
hampered due to lack of proper materials with large band gap, stable structure and easy fabrication.~\cite{MRS:9383312} 
In this work, based on first-principles calculations, we find monolayer of WX$_2$ and MoX$_2$ Haeckelite are 2D TIs
and the largest band gap is around 54 meV. Distinguished from other predicted QSH materials\cite{liu2011, PhysRevLett.111.136804, PhysRevB.89.115429, 2014arXiv1402.2399S, Zhou2014} based on honeycomb lattice, these MX$_2$-4-8 Haeckelites have square-like lattice and
a simple tight-binding model with one orbital per site and four sites per unit cell has been established
to achieve topologically nontrivial QSH state. Such extension from honeycomb lattice to square-like
lattice have largely broad the potential candidates for topological materials.~\cite{weng2014, XiangSQ}


The paper is arranged as follows. In section II we will introduce the details of first-principles
calculations. In section III, the calculation results are presented and TB analysis is performed. Finally,
section IV contains a conclusion of this work.

%

\section{Calculation method and crystal structure}

\begin{figure}[htp]
\includegraphics[width=5.2in]{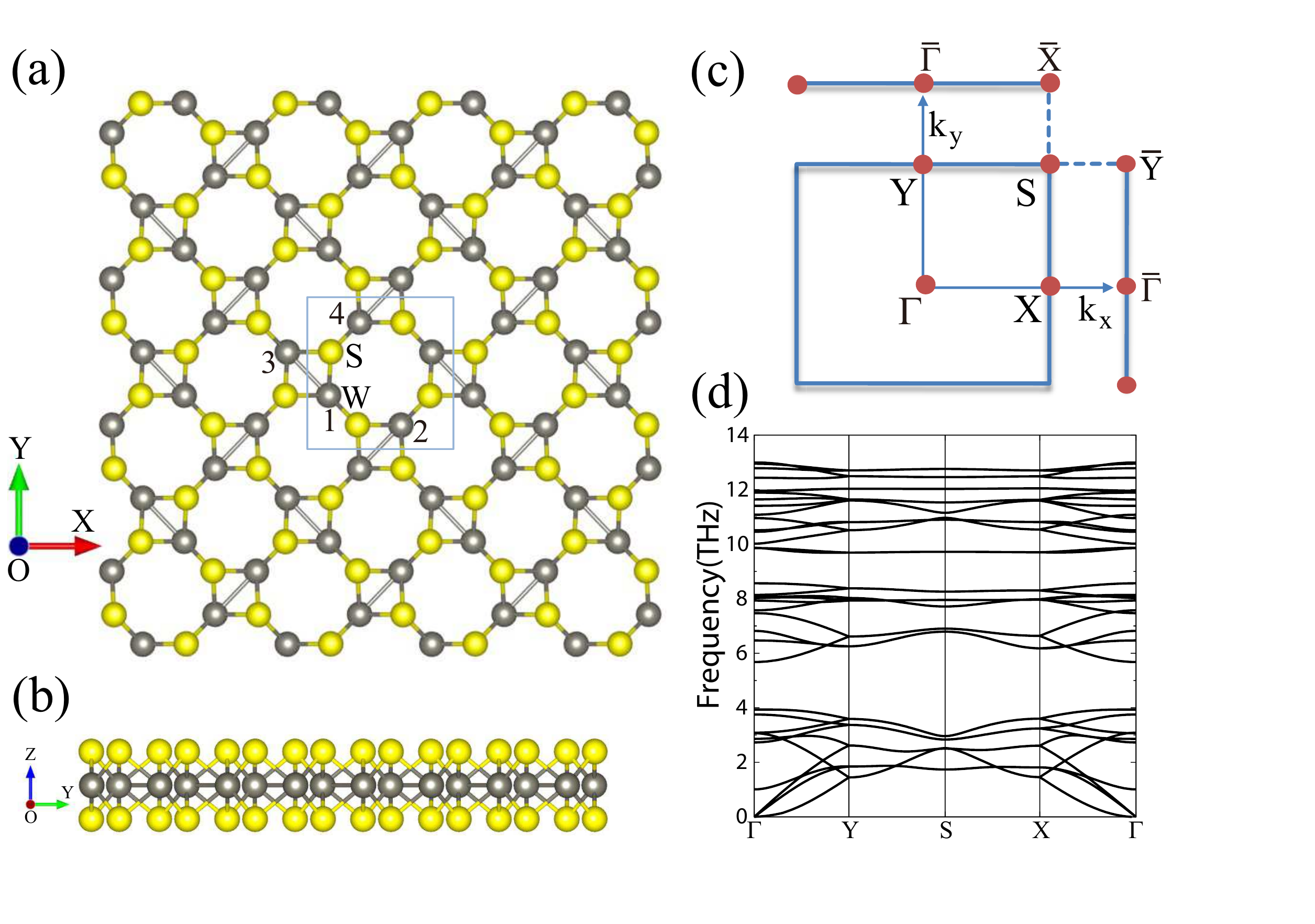}
      \caption{(color online). Top view(a) and side view (b) of relaxed WS$_2$-4-8.
      Gray ball is W and yellow ball is S. The primitive cell is shown in light blue
      rectangle. (c) 2D and projected edge first BZ with high symmetry point (red dots).
      (d) Phonon dispersion of WS$_2$-4-8.}
\label{structure}
\end{figure}

First-principles calculations were carried out using the projector
augmented wave method \cite{blochl1994,kresse1999} implemented in
Vienna $ab~ initio$ simulation package (VASP)\cite{kresse1996_1,kresse1996_2}.
Exchange and correlation potential was treated within the generalized gradient
approximation (GGA) of Perdew-Burke-Ernzerhof type\cite{Perdew1996}. SOC
was taken into account by the second variation method self-consistently. The cut-off energy
for plane wave expansion was 500 eV. The k-points sampling grid in the
self-consistent process was 9 $\times$ 9 $\times$ 3. The crystal structures
have been fully relaxed until the residual forces on each atom were less than
0.001 eV/\AA. The crystal parameters for all TMD Haeckelites
are shown in Table \ref{summary}. A vacuum of 20 \AA ~between layers was considered
in order to minimize image interactions from the periodic boundary condition.
PHONOPY has been employed to calculate the phonon dispersion \cite{togo2008}.
To explore the edge states of TMD Haeckelites, maximally localized wannier
functions (MLWFs) for the $d$ orbitals of W and $p$ orbitals of S have been constructed
and used to get $ab~ initio$ tight-binding (TB) hamiltonian\cite{marzari1997,souza2001,mostofi2008}.
Atomic SOC is added to the TB hamiltonian by fitting the first-principle calcualtions. The projected edge
states were obtained from the TB through an iterative method\cite{sancho1984quick,sancho1985highly, MRS:9383312}.

All TMD Haeckelites have the same non-symmorphic space group Pbam ($D_{2h}^{9}$). Except the difference of
lattice constants, all other TMD Haeckelites have nearly the same properties as WS$_2$. So we
choose WS$_2$-4-8 later as an example and the results of other TMD Haeckelites can be found in the Appendix. 
The relaxed crystal structure and Brillouin zone (BZ) for WS$_2$-4-8 are
shown in Fig. \ref{structure}. It is noted that bond length between W 1 and 3 is reduced compared to that in honeycomb lattice.
As we will see later, this bond is vital to topological phase transition.
The dynamic stability of WS$_2$-4-8 has been investigated by calculating it phonon spectrum. Imaginary frequencies can not be
found in the phonon dispersion of WS$_2$-4-8, which indicates the structure is stable (Fig. \ref{structure}(d)).

\begin{table}
\caption{Lattice parameters and band gaps for some TMD Haeckelites MX$_2$-4-8.}
\begin{tabular}{  l  c  c  c  c  c }
\hline
\hline
        Structure    & ~~~a(\AA)& & ~~~b(\AA)& ~~~Gap(meV) \\
\hline
        WS$_2$   & ~~~6.34  & & ~~~6.41  &~~~53.82 \\
        WSe$_2$  & ~~~6.40  & & ~~~6.86  &~~~30.03 \\
        WTe$_2$  & ~~~6.65  & & ~~~7.38  &~~~14.743\\
        MoS$_2$  & ~~~6.36  & & ~~~6.33  &~~~13.38 \\
        MoSe$_2$ & ~~~6.65  & & ~~~6.59  &~~~26.80 \\
\hline
\hline
\end{tabular}
\label{summary}
\end{table}

\section{results}

\begin{figure}[htp]
\includegraphics[width=5.2in]{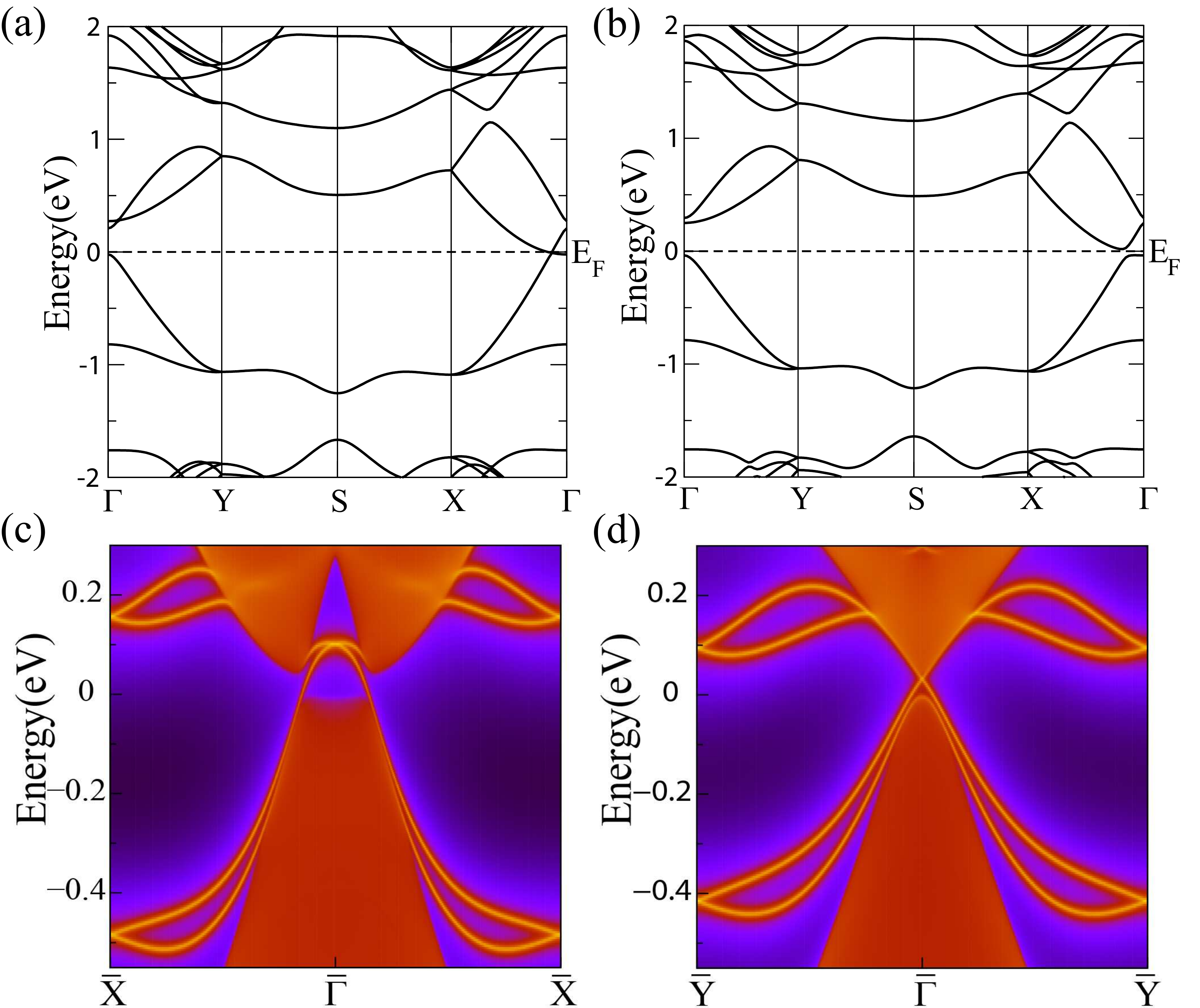}
      \caption{(color online). GGA (a) and GGA+SOC (b) band structures of WS$_2$-4-8.
      The calculated edge states for X (c) and Y (d) edge, respectively.}
      \label{bandstructure}
\end{figure}

Both GGA and GGA+SOC band structures of WS$_2$-4-8 are shown in Fig. 2.
Along $\Gamma$-X direction, there is a band crossing in GGA band structure.
The little group of k point located on this direction is $C_{2v}$. The two crossing
bands belong to different irreducible representations and such band crossing is protected by
\{$C_{2x}|\frac{1}{2}\frac{1}{2}0$\} operation. When SOC is included, as we can see in 
Fig.\ref{bandstructure} (b), WS$_2$-4-8 is a well defined insulator with indirect
band gap around 53.8 meV. Gaps for other TMD Haeckelites are listed in the
Table \ref{summary}. For system possesses both time reversal (TR) and space
inversion symmetry, the parity criterion proposed by Fu and Kane is a
convenient method to judge its band topology $Z_2$ number.~\cite{Fu:2007ei} Since the
space group of it is non-symmorphic, similar as that in single layer ZrTe$_5$ and HfTe$_5$~\cite{weng2014}, all
the bands at the three time reversal invariant momenta X, S, Y having degeneracy of even and odd states.
Only the band inversion at $\Gamma$ can leads to nontrivial band topology. It is true that the total
parity of occupied states at $\Gamma$ is -1 and the other three are +1. Therefore, the topological 
index $Z_2$ equals to 1, which means WS$_2$-4-8 is a QSH insulator. Considering the possible
underestimation of band gap of GGA, non-local Heyd-Scuseria-Ernzerhof (HSE06)
hybrid functional\cite{heyd2003hybrid,*heyd2006hybrid} is further supplemented
to check the topological property. The band topology does not change.

Since the existence of nontrivial edge states is the hallmark of QSH effect,
we have calculated the edge states of WS$_2$-4-8. As shown in Fig. \ref{bandstructure} (c, d),
there is edge Dirac cone dispersion connecting the bulk occupied and unoccupied states with 
Dirac point at $\bar{\Gamma}$ for both X and Y edges.

In order to understand the band inversion process at $\Gamma$ point explicitly,
a Slater-Koster TB has been constructed. Obviously, the main physics comes from the isolated group
of four bands around the Fermi level. The band character and projected
density of states (PDOS) analyses indicate that the low energy bands near the
fermi level are mainly contributed by the $d_{z^2}$ and $d_{x^2-y^2}$ orbitals of W.
The $d_{x^2-y^2}$ can mix with $d_{z^2}$ due to the distortion of 4-8 defects. For simplicity, only 
$d_{z^2}$ is taken into account for each W and one primitive cell contains four W atoms.
Therefore, it is possible and reasonable to construct a $4\times4$ TB hamiltonian for non-SOC case.
The non-SOC TB hamiltonian with four localized $d_{z^2}$ bases $|\omega_i\rangle$ (i=1, 2, 3, 4) can 
be written as follows in momentum space

  \begin{widetext}
  \begin{eqnarray}
        H_0(\boldsymbol{k}) = \left( \begin{array}{cccc}
            0 & 2t_1\cos(\frac{1}{2}k_x) & t_2 e^{\frac{i}{2}(-k_x+k_y)} &
                2t_3\cos(\frac{1}{2}k_y) \\
            2t_1\cos(\frac{1}{2}k_x) & 0 & 2t_3\cos(\frac{1}{2}k_y) &
                t_2 e^{\frac{i}{2}(-k_x-k_y)} \\
            t_2 e^{\frac{i}{2}(k_x-k_y)} & 2t_3\cos(\frac{1}{2}k_y) &
                0 & 2t_1\cos(\frac{1}{2}k_x) \\
            2t_3\cos(\frac{1}{2}k_y) & t_2 e^{\frac{i}{2}(k_x+k_y)} &
                 2t_1\cos(\frac{1}{2}k_x) & 0
         \end{array}\right)
         \label{H0k} \\ \nonumber
    \end{eqnarray}
   \end{widetext}

   where the hopping parameters are defined as
     \begin{eqnarray}
        t_1 = \langle w_1 | H_0 | w_2 \rangle \\
        t_2 = \langle w_1 | H_0 | w_{3} \rangle \\
        t_3 = \langle w_1 | H_0 | w_4 \rangle.
    \end{eqnarray}

    $|\omega_i\rangle$ means the orbital $d_{z^2}$ on $i$-th W atom labelled in Fig. \ref{structure} (a).
    However, one should note that periodic boundary condition is sacrificed here in order to get
    a concise form of Eq. (\ref{H0k}). When the calculated  properties concern global phase
    such as berry phase, we need to transform $H_0(\boldsymbol{k})$ into another form which
    satisfies periodic boundary condition.

   Similar to graphene, SOC is a second order effect for WS$_2$-4-8. The intrinsic atomic SOC for W is of
   the order of 200 meV while it is about 4 meV for graphene. Compare to the extremely small gap
   for graphene,  a large band gap (53.8 meV) is obtained for WS$_2$-4-8 at last. The hybridization
   between $|\omega_i\uparrow\rangle$ and $|\omega_j\downarrow\rangle$ is zero due to $m_z$ symmetry.
   If TR symmetry is preserved, we will
   have $H_{so}^{\uparrow\uparrow}(\boldsymbol{k})=H_{so}^{\downarrow\downarrow}(\boldsymbol{k})^T$.
   Therefore, even the hamiltonian size will be doubled when SOC is taken into account, we can still focus on
   spin up (spin down) subspace only. Spin down (spin up) subspace can be obtained using above
   restricted conditions. Considering all the symmetries, we obtain a generic matrix form for
   $H_{so}^{\uparrow\uparrow}(\boldsymbol{k})$.

  \begin{widetext}
    \begin{eqnarray}
        H_{so}^{\uparrow\uparrow}(\boldsymbol{k}) = \left(\begin{array}{cccc}
            0 & 2\lambda_1\cos(\frac{1}{2}k_x) & 0 & 2\lambda_3\cos(\frac{1}{2}k_y) \\
            2\lambda_1^*\cos(\frac{1}{2}k_x) & 0 & 2\lambda_3^*\cos(\frac{1}{2}k_y) & 0 \\
            0 & 2\lambda_3\cos(\frac{1}{2}k_y) & 0 & 2\lambda_1\cos(\frac{1}{2}k_x) \\
            2\lambda_3^*\cos(\frac{1}{2}k_y) & 0 & 2\lambda_1^*\cos(\frac{1}{2}k_x) & 0
            \end{array}\right)  \label{Hsok}
    \end{eqnarray}
    \end{widetext}

    where $\lambda_1$ and $\lambda_3 $ (pure imaginary numbers) are defined as
    \begin{align}
        \lambda_1 &= \langle w_1\uparrow | H_{so} | w_2\uparrow\rangle \\
        \lambda_3 &= \langle w_1\uparrow | H_{so} | w_4\uparrow\rangle
    \end{align}

\begin{figure}[htp]
\includegraphics[width=5.2in]{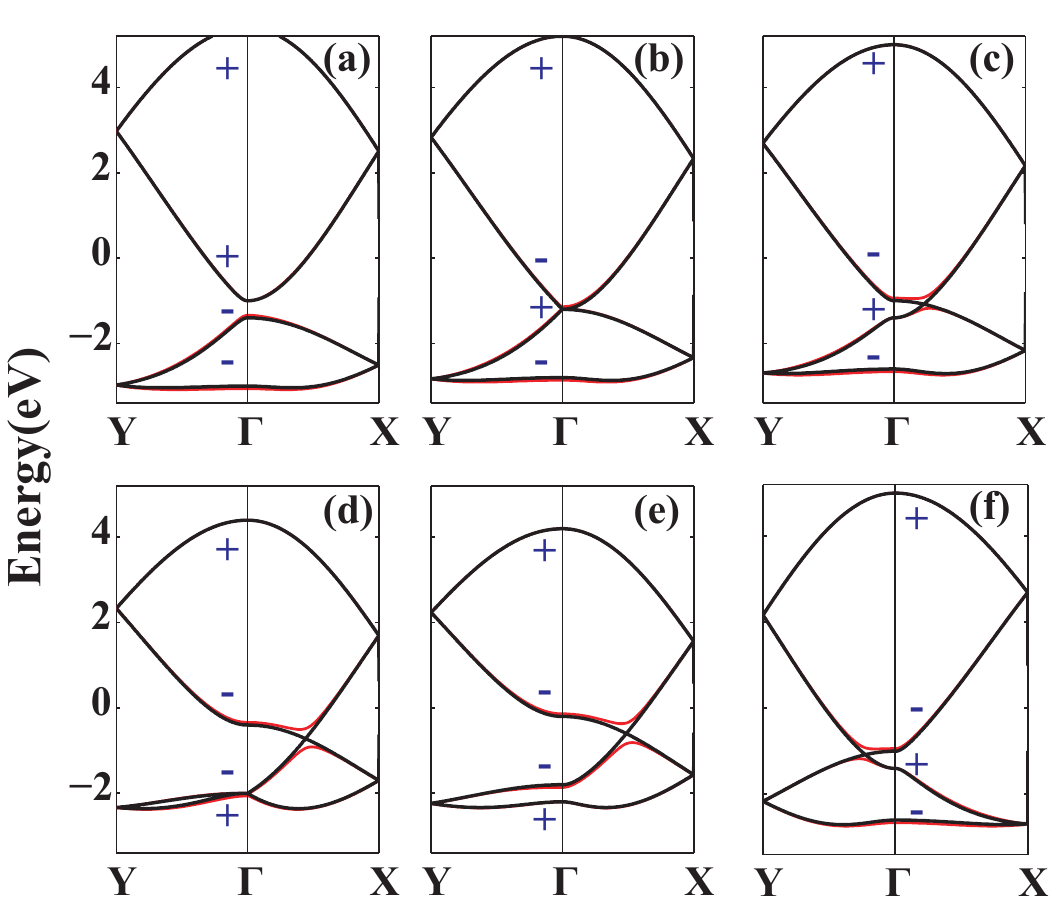}
      \caption{(color online). Non-SOC (black lines) and SOC (red lines) band structure evolution
      calculated with different $t_2$ in TB model. The parameters are $\lambda_1=-\lambda_3=0.08~i$,
      $t_1=1$,  $t_3=0.6$, $t_2=2.2$ (a), $t_2=2.0$ (b), $t_2=1.8$ (c),  $t_2=1.2$ (d) and $t_2=1.0$ (e).
      (f) $t_1=0.6$,  $t_3=1$ and $t_2=1.8$. Parity information at $\Gamma$ has been given.}
      \label{TBband}
\end{figure}

    
    Only the total sign($t_1t_2t_3$) makes sense. Sign$(t_1 t_2 t_3)<0$ can be inferred by inverting the
    whole bands of sign$(t_1 t_2 t_3)>0$. So we take $t_1>0$, $t_2>0$ and $t_3>0$. As discussed above, 
    only the band inversion at $\Gamma$ point can change the band topology. The four eigen states and their parities can be obtained explicitly as following:

    \begin{align}
    E_1 &= 2 t1 + t2 + 2 t3   ~~~~~~\text{  ~~parity~~ +}\\
    E_2 &= -2 t1 + t2 - 2 t3  ~~~~ \text{   ~~parity~~ +}\\
    E_3 &= 2 t1 - t2 - 2 t3   ~~~~~~\text{  ~~parity~~ }-\\
    E_4 &= -2 t1 - t2 + 2 t3  ~~~~\text{~~  parity~~ }-
      \end{align}

   Obviously, the $E_1$ has the highest energy. The band inversion happens between $E_2$ and $E_3$ or $E_2$ and $E_4$ can
   lead to QSH state. That means the band inversion will exist as long as $|t_2|$$<$max($|2t_1|$, $|2t_3|$). Fig. \ref{TBband} show
   the band structure calculated with the TB model with different sets of parameters. When $|t_2|>2|t_1|>2|t_3|$ (Fig. \ref{TBband} (a)), 
   the bonding and anti-bonding states are far away from each other. This is a trivial insulator. There will be a band touching at $\Gamma$
   point if $|t_2|=2|t_1|$ (Fig. \ref{TBband} (b)). When $2|t_1|>|t_2|>2|t_3|$, band inversion will
   happen (Fig. \ref{TBband} (c)) and the system enters into QSH state. The two valence bands ($E_2$ and $E_4$) with opposite parity 
   will be degenerate at $\Gamma$ if $|t_2|=2|t_3|$ (Fig. \ref{TBband} (d)). When $2|t_1|>2|t_3|> |t_2|$, they will separate
   each other (Fig. \ref{TBband} (e)). Topological phase transition will not take place in this 
   parametric region from Fig. \ref{TBband}(c) to (e), since the band inversion process occurs between two occupied valence bands.
   Another interesting thing is that $t_1$ and $t_3$ control the position of band crossing in non-SOC band structure. If $t_1>t_3$, the crossing 
   will be situated at $\Gamma-\text{X}$ direction (Fig. \ref{TBband} (c)) and it will move to $\Gamma-\text{Y}$ direction (Mo case in Appendix) 
   for $t_1<t_3$ (Fig. \ref{TBband} (f)). 

    Uniaxial strain effect has also been investigated with first-principles calculations, which can be used to tune the relative strength of these 
    hopping parameters and control the topological quantum state transition. If we fix crystal constant $a$ and decrease $b$, $t_3$ will increase 
    faster than $t_1$ and $t_2$. This means the band crossing in non-SOC band structure will move from $\Gamma-\text{X}$ to $\Gamma-\text{Y}$
    and uniaxial strain is always beneficial for band inversion. Similar results can be get for uniaxial strain along b-axis. Therefore the topology in 
    WS$_2$-4-8 is robust against uniaxial strain.

\section{CONCLUSION}

In summary, we have performed first-principles calculations for the electronic properties of TMD Haeckelites MX$_2$-4-8 and found
that they are 2D TIs. A simple TB model with one orbital per site and four sites per unit cell has been established to understand the 
band inversion mechanism. The nearest hopping parameter $t_2$ is vital to trigger the topological phase transition and can be tuned through
lattice strain effect. Such simple square-like lattice model to achieve various topological quantum states is very stimulating. It will lead further
study based on square lattice instead of honeycomb lattice and largely extend the searching range for topological materials.

\begin{acknowledgments}
We acknowledge the supports from National Natural Science Foundation of China (No. 11274359 and 11422428), 
the 973 program of China (No. 2011CBA00108 and 2013CB921700) and the ``Strategic Priority Research Program (B)" 
of the Chinese Academy of Sciences (No. XDB07020100). 
\end{acknowledgments}

\bibliography{haeckelite}

\begin{thebibliography}{48}%
\makeatletter
\providecommand \@ifxundefined [1]{%
 \@ifx{#1\undefined}
}%
\providecommand \@ifnum [1]{%
 \ifnum #1\expandafter \@firstoftwo
 \else \expandafter \@secondoftwo
 \fi
}%
\providecommand \@ifx [1]{%
 \ifx #1\expandafter \@firstoftwo
 \else \expandafter \@secondoftwo
 \fi
}%
\providecommand \natexlab [1]{#1}%
\providecommand \enquote  [1]{``#1''}%
\providecommand \bibnamefont  [1]{#1}%
\providecommand \bibfnamefont [1]{#1}%
\providecommand \citenamefont [1]{#1}%
\providecommand \href@noop [0]{\@secondoftwo}%
\providecommand \href [0]{\begingroup \@sanitize@url \@href}%
\providecommand \@href[1]{\@@startlink{#1}\@@href}%
\providecommand \@@href[1]{\endgroup#1\@@endlink}%
\providecommand \@sanitize@url [0]{\catcode `\\12\catcode `\$12\catcode
  `\&12\catcode `\#12\catcode `\^12\catcode `\_12\catcode `\%12\relax}%
\providecommand \@@startlink[1]{}%
\providecommand \@@endlink[0]{}%
\providecommand \url  [0]{\begingroup\@sanitize@url \@url }%
\providecommand \@url [1]{\endgroup\@href {#1}{\urlprefix }}%
\providecommand \urlprefix  [0]{URL }%
\providecommand \Eprint [0]{\href }%
\providecommand \doibase [0]{http://dx.doi.org/}%
\providecommand \selectlanguage [0]{\@gobble}%
\providecommand \bibinfo  [0]{\@secondoftwo}%
\providecommand \bibfield  [0]{\@secondoftwo}%
\providecommand \translation [1]{[#1]}%
\providecommand \BibitemOpen [0]{}%
\providecommand \bibitemStop [0]{}%
\providecommand \bibitemNoStop [0]{.\EOS\space}%
\providecommand \EOS [0]{\spacefactor3000\relax}%
\providecommand \BibitemShut  [1]{\csname bibitem#1\endcsname}%
\let\auto@bib@innerbib\@empty
\bibitem [{\citenamefont {Novoselov}\ \emph {et~al.}(2004)\citenamefont
  {Novoselov}, \citenamefont {Geim}, \citenamefont {Morozov}, \citenamefont
  {Jiang}, \citenamefont {Zhang}, \citenamefont {Dubonos}, \citenamefont
  {Grigorieva},\ and\ \citenamefont {Firsov}}]{novoselov2004}%
  \BibitemOpen
  \bibfield  {author} {\bibinfo {author} {\bibfnamefont {K.~S.}\ \bibnamefont
  {Novoselov}}, \bibinfo {author} {\bibfnamefont {A.~K.}\ \bibnamefont {Geim}},
  \bibinfo {author} {\bibfnamefont {S.~V.}\ \bibnamefont {Morozov}}, \bibinfo
  {author} {\bibfnamefont {D.}~\bibnamefont {Jiang}}, \bibinfo {author}
  {\bibfnamefont {Y.}~\bibnamefont {Zhang}}, \bibinfo {author} {\bibfnamefont
  {S.~V.}\ \bibnamefont {Dubonos}}, \bibinfo {author} {\bibfnamefont {I.~V.}\
  \bibnamefont {Grigorieva}}, \ and\ \bibinfo {author} {\bibfnamefont {A.~A.}\
  \bibnamefont {Firsov}},\ }\href@noop {} {\bibfield  {journal} {\bibinfo
  {journal} {science}\ }\textbf {\bibinfo {volume} {306}},\ \bibinfo {pages}
  {666} (\bibinfo {year} {2004})}\BibitemShut {NoStop}%
\bibitem [{\citenamefont {Radisavljevic}\ and\ \citenamefont
  {Kis}(2013)}]{radisavljevic2013mobility}%
  \BibitemOpen
  \bibfield  {author} {\bibinfo {author} {\bibfnamefont {B.}~\bibnamefont
  {Radisavljevic}}\ and\ \bibinfo {author} {\bibfnamefont {A.}~\bibnamefont
  {Kis}},\ }\href@noop {} {\bibfield  {journal} {\bibinfo  {journal} {Nature
  materials}\ }\textbf {\bibinfo {volume} {12}},\ \bibinfo {pages} {815}
  (\bibinfo {year} {2013})}\BibitemShut {NoStop}%
\bibitem [{\citenamefont {Wang}\ \emph {et~al.}(2012)\citenamefont {Wang},
  \citenamefont {Kalantar-Zadeh}, \citenamefont {Kis}, \citenamefont
  {Coleman},\ and\ \citenamefont {Strano}}]{wang2012electronics}%
  \BibitemOpen
  \bibfield  {author} {\bibinfo {author} {\bibfnamefont {Q.~H.}\ \bibnamefont
  {Wang}}, \bibinfo {author} {\bibfnamefont {K.}~\bibnamefont
  {Kalantar-Zadeh}}, \bibinfo {author} {\bibfnamefont {A.}~\bibnamefont {Kis}},
  \bibinfo {author} {\bibfnamefont {J.~N.}\ \bibnamefont {Coleman}}, \ and\
  \bibinfo {author} {\bibfnamefont {M.~S.}\ \bibnamefont {Strano}},\
  }\href@noop {} {\bibfield  {journal} {\bibinfo  {journal} {Nature
  nanotechnology}\ }\textbf {\bibinfo {volume} {7}},\ \bibinfo {pages} {699}
  (\bibinfo {year} {2012})}\BibitemShut {NoStop}%
\bibitem [{\citenamefont {Xu}\ \emph {et~al.}(2014)\citenamefont {Xu},
  \citenamefont {Wang}, \citenamefont {Yan},\ and\ \citenamefont
  {Qi}}]{xu2014topological}%
  \BibitemOpen
  \bibfield  {author} {\bibinfo {author} {\bibfnamefont {G.}~\bibnamefont
  {Xu}}, \bibinfo {author} {\bibfnamefont {J.}~\bibnamefont {Wang}}, \bibinfo
  {author} {\bibfnamefont {B.}~\bibnamefont {Yan}}, \ and\ \bibinfo {author}
  {\bibfnamefont {X.-L.}\ \bibnamefont {Qi}},\ }\href@noop {} {\bibfield
  {journal} {\bibinfo  {journal} {Physical Review B}\ }\textbf {\bibinfo
  {volume} {90}},\ \bibinfo {pages} {100505} (\bibinfo {year}
  {2014})}\BibitemShut {NoStop}%
\bibitem [{\citenamefont {Shirodkar}\ and\ \citenamefont
  {Waghmare}(2014)}]{shirodkar2014emergence}%
  \BibitemOpen
  \bibfield  {author} {\bibinfo {author} {\bibfnamefont {S.~N.}\ \bibnamefont
  {Shirodkar}}\ and\ \bibinfo {author} {\bibfnamefont {U.~V.}\ \bibnamefont
  {Waghmare}},\ }\href@noop {} {\bibfield  {journal} {\bibinfo  {journal}
  {Physical review letters}\ }\textbf {\bibinfo {volume} {112}},\ \bibinfo
  {pages} {157601} (\bibinfo {year} {2014})}\BibitemShut {NoStop}%
\bibitem [{\citenamefont {Cao}\ \emph {et~al.}(2012)\citenamefont {Cao},
  \citenamefont {Wang}, \citenamefont {Han}, \citenamefont {Ye}, \citenamefont
  {Zhu}, \citenamefont {Shi}, \citenamefont {Niu}, \citenamefont {Tan},
  \citenamefont {Wang}, \citenamefont {Liu} \emph {et~al.}}]{cao2012valley}%
  \BibitemOpen
  \bibfield  {author} {\bibinfo {author} {\bibfnamefont {T.}~\bibnamefont
  {Cao}}, \bibinfo {author} {\bibfnamefont {G.}~\bibnamefont {Wang}}, \bibinfo
  {author} {\bibfnamefont {W.}~\bibnamefont {Han}}, \bibinfo {author}
  {\bibfnamefont {H.}~\bibnamefont {Ye}}, \bibinfo {author} {\bibfnamefont
  {C.}~\bibnamefont {Zhu}}, \bibinfo {author} {\bibfnamefont {J.}~\bibnamefont
  {Shi}}, \bibinfo {author} {\bibfnamefont {Q.}~\bibnamefont {Niu}}, \bibinfo
  {author} {\bibfnamefont {P.}~\bibnamefont {Tan}}, \bibinfo {author}
  {\bibfnamefont {E.}~\bibnamefont {Wang}}, \bibinfo {author} {\bibfnamefont
  {B.}~\bibnamefont {Liu}},  \emph {et~al.},\ }\href@noop {} {\bibfield
  {journal} {\bibinfo  {journal} {Nature communications}\ }\textbf {\bibinfo
  {volume} {3}},\ \bibinfo {pages} {887} (\bibinfo {year} {2012})}\BibitemShut
  {NoStop}%
\bibitem [{\citenamefont {Novoselov}\ \emph {et~al.}(2005)\citenamefont
  {Novoselov}, \citenamefont {Jiang}, \citenamefont {Schedin}, \citenamefont
  {Booth}, \citenamefont {Khotkevich}, \citenamefont {Morozov},\ and\
  \citenamefont {Geim}}]{novoselov2005}%
  \BibitemOpen
  \bibfield  {author} {\bibinfo {author} {\bibfnamefont {K.~S.}\ \bibnamefont
  {Novoselov}}, \bibinfo {author} {\bibfnamefont {D.}~\bibnamefont {Jiang}},
  \bibinfo {author} {\bibfnamefont {F.}~\bibnamefont {Schedin}}, \bibinfo
  {author} {\bibfnamefont {T.~J.}\ \bibnamefont {Booth}}, \bibinfo {author}
  {\bibfnamefont {V.~V.}\ \bibnamefont {Khotkevich}}, \bibinfo {author}
  {\bibfnamefont {S.~V.}\ \bibnamefont {Morozov}}, \ and\ \bibinfo {author}
  {\bibfnamefont {A.~K.}\ \bibnamefont {Geim}},\ }\href@noop {} {\bibfield
  {journal} {\bibinfo  {journal} {Proceedings of the National Academy of
  Sciences of the United States of America}\ }\textbf {\bibinfo {volume}
  {102}},\ \bibinfo {pages} {10451} (\bibinfo {year} {2005})}\BibitemShut
  {NoStop}%
\bibitem [{\citenamefont {Radisavljevic}\ \emph {et~al.}(2011)\citenamefont
  {Radisavljevic}, \citenamefont {Radenovic}, \citenamefont {Brivio},
  \citenamefont {Giacometti},\ and\ \citenamefont {Kis}}]{radisavljevic2011}%
  \BibitemOpen
  \bibfield  {author} {\bibinfo {author} {\bibfnamefont {B.}~\bibnamefont
  {Radisavljevic}}, \bibinfo {author} {\bibfnamefont {A.}~\bibnamefont
  {Radenovic}}, \bibinfo {author} {\bibfnamefont {J.}~\bibnamefont {Brivio}},
  \bibinfo {author} {\bibfnamefont {V.}~\bibnamefont {Giacometti}}, \ and\
  \bibinfo {author} {\bibfnamefont {A.}~\bibnamefont {Kis}},\ }\href@noop {}
  {\bibfield  {journal} {\bibinfo  {journal} {Nature nanotechnology}\ }\textbf
  {\bibinfo {volume} {6}},\ \bibinfo {pages} {147} (\bibinfo {year}
  {2011})}\BibitemShut {NoStop}%
\bibitem [{\citenamefont {Eda}\ \emph {et~al.}(2011)\citenamefont {Eda},
  \citenamefont {Yamaguchi}, \citenamefont {Voiry}, \citenamefont {Fujita},
  \citenamefont {Chen},\ and\ \citenamefont {Chhowalla}}]{eda2011}%
  \BibitemOpen
  \bibfield  {author} {\bibinfo {author} {\bibfnamefont {G.}~\bibnamefont
  {Eda}}, \bibinfo {author} {\bibfnamefont {H.}~\bibnamefont {Yamaguchi}},
  \bibinfo {author} {\bibfnamefont {D.}~\bibnamefont {Voiry}}, \bibinfo
  {author} {\bibfnamefont {T.}~\bibnamefont {Fujita}}, \bibinfo {author}
  {\bibfnamefont {M.}~\bibnamefont {Chen}}, \ and\ \bibinfo {author}
  {\bibfnamefont {M.}~\bibnamefont {Chhowalla}},\ }\href@noop {} {\bibfield
  {journal} {\bibinfo  {journal} {Nano letters}\ }\textbf {\bibinfo {volume}
  {11}},\ \bibinfo {pages} {5111} (\bibinfo {year} {2011})}\BibitemShut
  {NoStop}%
\bibitem [{\citenamefont {Coleman}\ \emph {et~al.}(2011)\citenamefont
  {Coleman}, \citenamefont {Lotya}, \citenamefont {O'Neill}, \citenamefont
  {Bergin}, \citenamefont {King}, \citenamefont {Khan}, \citenamefont {Young},
  \citenamefont {Gaucher}, \citenamefont {De}, \citenamefont {Smith} \emph
  {et~al.}}]{coleman2011}%
  \BibitemOpen
  \bibfield  {author} {\bibinfo {author} {\bibfnamefont {J.~N.}\ \bibnamefont
  {Coleman}}, \bibinfo {author} {\bibfnamefont {M.}~\bibnamefont {Lotya}},
  \bibinfo {author} {\bibfnamefont {A.}~\bibnamefont {O'Neill}}, \bibinfo
  {author} {\bibfnamefont {S.~D.}\ \bibnamefont {Bergin}}, \bibinfo {author}
  {\bibfnamefont {P.~J.}\ \bibnamefont {King}}, \bibinfo {author}
  {\bibfnamefont {U.}~\bibnamefont {Khan}}, \bibinfo {author} {\bibfnamefont
  {K.}~\bibnamefont {Young}}, \bibinfo {author} {\bibfnamefont
  {A.}~\bibnamefont {Gaucher}}, \bibinfo {author} {\bibfnamefont
  {S.}~\bibnamefont {De}}, \bibinfo {author} {\bibfnamefont {R.~J.}\
  \bibnamefont {Smith}},  \emph {et~al.},\ }\href@noop {} {\bibfield  {journal}
  {\bibinfo  {journal} {Science}\ }\textbf {\bibinfo {volume} {331}},\ \bibinfo
  {pages} {568} (\bibinfo {year} {2011})}\BibitemShut {NoStop}%
\bibitem [{\citenamefont {Lee}\ \emph {et~al.}(2012)\citenamefont {Lee},
  \citenamefont {Zhang}, \citenamefont {Zhang}, \citenamefont {Chang},
  \citenamefont {Lin}, \citenamefont {Chang}, \citenamefont {Yu}, \citenamefont
  {Wang}, \citenamefont {Chang}, \citenamefont {Li} \emph {et~al.}}]{lee2012}%
  \BibitemOpen
  \bibfield  {author} {\bibinfo {author} {\bibfnamefont {Y.-H.}\ \bibnamefont
  {Lee}}, \bibinfo {author} {\bibfnamefont {X.-Q.}\ \bibnamefont {Zhang}},
  \bibinfo {author} {\bibfnamefont {W.}~\bibnamefont {Zhang}}, \bibinfo
  {author} {\bibfnamefont {M.-T.}\ \bibnamefont {Chang}}, \bibinfo {author}
  {\bibfnamefont {C.-T.}\ \bibnamefont {Lin}}, \bibinfo {author} {\bibfnamefont
  {K.-D.}\ \bibnamefont {Chang}}, \bibinfo {author} {\bibfnamefont {Y.-C.}\
  \bibnamefont {Yu}}, \bibinfo {author} {\bibfnamefont {J.~T.-W.}\ \bibnamefont
  {Wang}}, \bibinfo {author} {\bibfnamefont {C.-S.}\ \bibnamefont {Chang}},
  \bibinfo {author} {\bibfnamefont {L.-J.}\ \bibnamefont {Li}},  \emph
  {et~al.},\ }\href@noop {} {\bibfield  {journal} {\bibinfo  {journal}
  {Advanced Materials}\ }\textbf {\bibinfo {volume} {24}},\ \bibinfo {pages}
  {2320} (\bibinfo {year} {2012})}\BibitemShut {NoStop}%
\bibitem [{\citenamefont {Liu}\ \emph {et~al.}(2012)\citenamefont {Liu},
  \citenamefont {Zhang}, \citenamefont {Lee}, \citenamefont {Lin},
  \citenamefont {Chang}, \citenamefont {Su}, \citenamefont {Chang},
  \citenamefont {Li}, \citenamefont {Shi}, \citenamefont {Zhang} \emph
  {et~al.}}]{liu2012}%
  \BibitemOpen
  \bibfield  {author} {\bibinfo {author} {\bibfnamefont {K.-K.}\ \bibnamefont
  {Liu}}, \bibinfo {author} {\bibfnamefont {W.}~\bibnamefont {Zhang}}, \bibinfo
  {author} {\bibfnamefont {Y.-H.}\ \bibnamefont {Lee}}, \bibinfo {author}
  {\bibfnamefont {Y.-C.}\ \bibnamefont {Lin}}, \bibinfo {author} {\bibfnamefont
  {M.-T.}\ \bibnamefont {Chang}}, \bibinfo {author} {\bibfnamefont {C.-Y.}\
  \bibnamefont {Su}}, \bibinfo {author} {\bibfnamefont {C.-S.}\ \bibnamefont
  {Chang}}, \bibinfo {author} {\bibfnamefont {H.}~\bibnamefont {Li}}, \bibinfo
  {author} {\bibfnamefont {Y.}~\bibnamefont {Shi}}, \bibinfo {author}
  {\bibfnamefont {H.}~\bibnamefont {Zhang}},  \emph {et~al.},\ }\href@noop {}
  {\bibfield  {journal} {\bibinfo  {journal} {Nano letters}\ }\textbf {\bibinfo
  {volume} {12}},\ \bibinfo {pages} {1538} (\bibinfo {year}
  {2012})}\BibitemShut {NoStop}%
\bibitem [{\citenamefont {Crespi}\ \emph {et~al.}(1996)\citenamefont {Crespi},
  \citenamefont {Benedict}, \citenamefont {Cohen},\ and\ \citenamefont
  {Louie}}]{crespi1996}%
  \BibitemOpen
  \bibfield  {author} {\bibinfo {author} {\bibfnamefont {V.~H.}\ \bibnamefont
  {Crespi}}, \bibinfo {author} {\bibfnamefont {L.~X.}\ \bibnamefont
  {Benedict}}, \bibinfo {author} {\bibfnamefont {M.~L.}\ \bibnamefont {Cohen}},
  \ and\ \bibinfo {author} {\bibfnamefont {S.~G.}\ \bibnamefont {Louie}},\
  }\href@noop {} {\bibfield  {journal} {\bibinfo  {journal} {Physical Review
  B}\ }\textbf {\bibinfo {volume} {53}},\ \bibinfo {pages} {R13303} (\bibinfo
  {year} {1996})}\BibitemShut {NoStop}%
\bibitem [{\citenamefont {Terrones}\ \emph {et~al.}(2000)\citenamefont
  {Terrones}, \citenamefont {Terrones}, \citenamefont {Hern{\'a}ndez},
  \citenamefont {Grobert}, \citenamefont {Charlier},\ and\ \citenamefont
  {Ajayan}}]{terrones2000}%
  \BibitemOpen
  \bibfield  {author} {\bibinfo {author} {\bibfnamefont {H.}~\bibnamefont
  {Terrones}}, \bibinfo {author} {\bibfnamefont {M.}~\bibnamefont {Terrones}},
  \bibinfo {author} {\bibfnamefont {E.}~\bibnamefont {Hern{\'a}ndez}}, \bibinfo
  {author} {\bibfnamefont {N.}~\bibnamefont {Grobert}}, \bibinfo {author}
  {\bibfnamefont {J.-C.}\ \bibnamefont {Charlier}}, \ and\ \bibinfo {author}
  {\bibfnamefont {P.}~\bibnamefont {Ajayan}},\ }\href@noop {} {\bibfield
  {journal} {\bibinfo  {journal} {Physical review letters}\ }\textbf {\bibinfo
  {volume} {84}},\ \bibinfo {pages} {1716} (\bibinfo {year}
  {2000})}\BibitemShut {NoStop}%
\bibitem [{\citenamefont {Sheng}\ \emph {et~al.}(2012)\citenamefont {Sheng},
  \citenamefont {Cui}, \citenamefont {Ye}, \citenamefont {Yan}, \citenamefont
  {Zheng},\ and\ \citenamefont {Su}}]{sheng2012octagraphene}%
  \BibitemOpen
  \bibfield  {author} {\bibinfo {author} {\bibfnamefont {X.-L.}\ \bibnamefont
  {Sheng}}, \bibinfo {author} {\bibfnamefont {H.-J.}\ \bibnamefont {Cui}},
  \bibinfo {author} {\bibfnamefont {F.}~\bibnamefont {Ye}}, \bibinfo {author}
  {\bibfnamefont {Q.-B.}\ \bibnamefont {Yan}}, \bibinfo {author} {\bibfnamefont
  {Q.-R.}\ \bibnamefont {Zheng}}, \ and\ \bibinfo {author} {\bibfnamefont
  {G.}~\bibnamefont {Su}},\ }\href@noop {} {\bibfield  {journal} {\bibinfo
  {journal} {Journal of Applied Physics}\ }\textbf {\bibinfo {volume} {112}},\
  \bibinfo {pages} {074315} (\bibinfo {year} {2012})}\BibitemShut {NoStop}%
\bibitem [{\citenamefont {Li}\ \emph {et~al.}(2014)\citenamefont {Li},
  \citenamefont {Guo}, \citenamefont {Zhang},\ and\ \citenamefont
  {Zhang}}]{PhysRevB.89.205402}%
  \BibitemOpen
  \bibfield  {author} {\bibinfo {author} {\bibfnamefont {W.}~\bibnamefont
  {Li}}, \bibinfo {author} {\bibfnamefont {M.}~\bibnamefont {Guo}}, \bibinfo
  {author} {\bibfnamefont {G.}~\bibnamefont {Zhang}}, \ and\ \bibinfo {author}
  {\bibfnamefont {Y.-W.}\ \bibnamefont {Zhang}},\ }\href {\doibase
  10.1103/PhysRevB.89.205402} {\bibfield  {journal} {\bibinfo  {journal} {Phys.
  Rev. B}\ }\textbf {\bibinfo {volume} {89}},\ \bibinfo {pages} {205402}
  (\bibinfo {year} {2014})}\BibitemShut {NoStop}%
\bibitem [{\citenamefont {Terrones}\ and\ \citenamefont
  {Terrones}(2014)}]{terrones2014}%
  \BibitemOpen
  \bibfield  {author} {\bibinfo {author} {\bibfnamefont {H.}~\bibnamefont
  {Terrones}}\ and\ \bibinfo {author} {\bibfnamefont {M.}~\bibnamefont
  {Terrones}},\ }\href@noop {} {\bibfield  {journal} {\bibinfo  {journal} {2D
  Materials}\ }\textbf {\bibinfo {volume} {1}},\ \bibinfo {pages} {011003}
  (\bibinfo {year} {2014})}\BibitemShut {NoStop}%
\bibitem [{\citenamefont {van~der Zande}\ \emph {et~al.}(2013)\citenamefont
  {van~der Zande}, \citenamefont {Huang}, \citenamefont {Chenet}, \citenamefont
  {Berkelbach}, \citenamefont {You}, \citenamefont {Lee}, \citenamefont
  {Heinz}, \citenamefont {Reichman}, \citenamefont {Muller},\ and\
  \citenamefont {Hone}}]{van2013}%
  \BibitemOpen
  \bibfield  {author} {\bibinfo {author} {\bibfnamefont {A.~M.}\ \bibnamefont
  {van~der Zande}}, \bibinfo {author} {\bibfnamefont {P.~Y.}\ \bibnamefont
  {Huang}}, \bibinfo {author} {\bibfnamefont {D.~A.}\ \bibnamefont {Chenet}},
  \bibinfo {author} {\bibfnamefont {T.~C.}\ \bibnamefont {Berkelbach}},
  \bibinfo {author} {\bibfnamefont {Y.}~\bibnamefont {You}}, \bibinfo {author}
  {\bibfnamefont {G.-H.}\ \bibnamefont {Lee}}, \bibinfo {author} {\bibfnamefont
  {T.~F.}\ \bibnamefont {Heinz}}, \bibinfo {author} {\bibfnamefont {D.~R.}\
  \bibnamefont {Reichman}}, \bibinfo {author} {\bibfnamefont {D.~A.}\
  \bibnamefont {Muller}}, \ and\ \bibinfo {author} {\bibfnamefont {J.~C.}\
  \bibnamefont {Hone}},\ }\href@noop {} {\bibfield  {journal} {\bibinfo
  {journal} {Nature materials}\ }\textbf {\bibinfo {volume} {12}},\ \bibinfo
  {pages} {554} (\bibinfo {year} {2013})}\BibitemShut {NoStop}%
\bibitem [{\citenamefont {Kotakoski}\ \emph {et~al.}(2011)\citenamefont
  {Kotakoski}, \citenamefont {Krasheninnikov}, \citenamefont {Kaiser},\ and\
  \citenamefont {Meyer}}]{kotakoski2011}%
  \BibitemOpen
  \bibfield  {author} {\bibinfo {author} {\bibfnamefont {J.}~\bibnamefont
  {Kotakoski}}, \bibinfo {author} {\bibfnamefont {A.}~\bibnamefont
  {Krasheninnikov}}, \bibinfo {author} {\bibfnamefont {U.}~\bibnamefont
  {Kaiser}}, \ and\ \bibinfo {author} {\bibfnamefont {J.}~\bibnamefont
  {Meyer}},\ }\href@noop {} {\bibfield  {journal} {\bibinfo  {journal}
  {Physical review letters}\ }\textbf {\bibinfo {volume} {106}},\ \bibinfo
  {pages} {105505} (\bibinfo {year} {2011})}\BibitemShut {NoStop}%
\bibitem [{\citenamefont {Komsa}\ \emph {et~al.}(2012)\citenamefont {Komsa},
  \citenamefont {Kotakoski}, \citenamefont {Kurasch}, \citenamefont {Lehtinen},
  \citenamefont {Kaiser},\ and\ \citenamefont {Krasheninnikov}}]{komsa2012}%
  \BibitemOpen
  \bibfield  {author} {\bibinfo {author} {\bibfnamefont {H.-P.}\ \bibnamefont
  {Komsa}}, \bibinfo {author} {\bibfnamefont {J.}~\bibnamefont {Kotakoski}},
  \bibinfo {author} {\bibfnamefont {S.}~\bibnamefont {Kurasch}}, \bibinfo
  {author} {\bibfnamefont {O.}~\bibnamefont {Lehtinen}}, \bibinfo {author}
  {\bibfnamefont {U.}~\bibnamefont {Kaiser}}, \ and\ \bibinfo {author}
  {\bibfnamefont {A.~V.}\ \bibnamefont {Krasheninnikov}},\ }\href@noop {}
  {\bibfield  {journal} {\bibinfo  {journal} {Physical review letters}\
  }\textbf {\bibinfo {volume} {109}},\ \bibinfo {pages} {035503} (\bibinfo
  {year} {2012})}\BibitemShut {NoStop}%
\bibitem [{\citenamefont {Qian}\ \emph {et~al.}(2014)\citenamefont {Qian},
  \citenamefont {Liu}, \citenamefont {Fu},\ and\ \citenamefont
  {Li}}]{qian2014quantum}%
  \BibitemOpen
  \bibfield  {author} {\bibinfo {author} {\bibfnamefont {X.}~\bibnamefont
  {Qian}}, \bibinfo {author} {\bibfnamefont {J.}~\bibnamefont {Liu}}, \bibinfo
  {author} {\bibfnamefont {L.}~\bibnamefont {Fu}}, \ and\ \bibinfo {author}
  {\bibfnamefont {J.}~\bibnamefont {Li}},\ }\href@noop {} {\bibfield  {journal}
  {\bibinfo  {journal} {Science}\ }\textbf {\bibinfo {volume} {346}},\ \bibinfo
  {pages} {1344} (\bibinfo {year} {2014})}\BibitemShut {NoStop}%
\bibitem [{\citenamefont {Kane}\ and\ \citenamefont {Mele}(2005)}]{kane2005}%
  \BibitemOpen
  \bibfield  {author} {\bibinfo {author} {\bibfnamefont {C.~L.}\ \bibnamefont
  {Kane}}\ and\ \bibinfo {author} {\bibfnamefont {E.~J.}\ \bibnamefont
  {Mele}},\ }\href@noop {} {\bibfield  {journal} {\bibinfo  {journal} {Physical
  Review Letters}\ }\textbf {\bibinfo {volume} {95}},\ \bibinfo {pages}
  {226801} (\bibinfo {year} {2005})}\BibitemShut {NoStop}%
\bibitem [{\citenamefont {Bernevig}\ \emph {et~al.}(2006)\citenamefont
  {Bernevig}, \citenamefont {Hughes},\ and\ \citenamefont
  {Zhang}}]{bernevig2006}%
  \BibitemOpen
  \bibfield  {author} {\bibinfo {author} {\bibfnamefont {B.~A.}\ \bibnamefont
  {Bernevig}}, \bibinfo {author} {\bibfnamefont {T.~L.}\ \bibnamefont
  {Hughes}}, \ and\ \bibinfo {author} {\bibfnamefont {S.-C.}\ \bibnamefont
  {Zhang}},\ }\href@noop {} {\bibfield  {journal} {\bibinfo  {journal}
  {Science}\ }\textbf {\bibinfo {volume} {314}},\ \bibinfo {pages} {1757}
  (\bibinfo {year} {2006})}\BibitemShut {NoStop}%
\bibitem [{\citenamefont {Yao}\ \emph {et~al.}(2007)\citenamefont {Yao},
  \citenamefont {Ye}, \citenamefont {Qi}, \citenamefont {Zhang},\ and\
  \citenamefont {Fang}}]{yao2007spin}%
  \BibitemOpen
  \bibfield  {author} {\bibinfo {author} {\bibfnamefont {Y.}~\bibnamefont
  {Yao}}, \bibinfo {author} {\bibfnamefont {F.}~\bibnamefont {Ye}}, \bibinfo
  {author} {\bibfnamefont {X.-L.}\ \bibnamefont {Qi}}, \bibinfo {author}
  {\bibfnamefont {S.-C.}\ \bibnamefont {Zhang}}, \ and\ \bibinfo {author}
  {\bibfnamefont {Z.}~\bibnamefont {Fang}},\ }\href@noop {} {\bibfield
  {journal} {\bibinfo  {journal} {Physical Review B}\ }\textbf {\bibinfo
  {volume} {75}},\ \bibinfo {pages} {041401} (\bibinfo {year}
  {2007})}\BibitemShut {NoStop}%
\bibitem [{\citenamefont {K{\"o}nig}\ \emph {et~al.}(2007)\citenamefont
  {K{\"o}nig}, \citenamefont {Wiedmann}, \citenamefont {Br{\"u}ne},
  \citenamefont {Roth}, \citenamefont {Buhmann}, \citenamefont {Molenkamp},
  \citenamefont {Qi},\ and\ \citenamefont {Zhang}}]{konig2007}%
  \BibitemOpen
  \bibfield  {author} {\bibinfo {author} {\bibfnamefont {M.}~\bibnamefont
  {K{\"o}nig}}, \bibinfo {author} {\bibfnamefont {S.}~\bibnamefont {Wiedmann}},
  \bibinfo {author} {\bibfnamefont {C.}~\bibnamefont {Br{\"u}ne}}, \bibinfo
  {author} {\bibfnamefont {A.}~\bibnamefont {Roth}}, \bibinfo {author}
  {\bibfnamefont {H.}~\bibnamefont {Buhmann}}, \bibinfo {author} {\bibfnamefont
  {L.~W.}\ \bibnamefont {Molenkamp}}, \bibinfo {author} {\bibfnamefont {X.-L.}\
  \bibnamefont {Qi}}, \ and\ \bibinfo {author} {\bibfnamefont {S.-C.}\
  \bibnamefont {Zhang}},\ }\href@noop {} {\bibfield  {journal} {\bibinfo
  {journal} {Science}\ }\textbf {\bibinfo {volume} {318}},\ \bibinfo {pages}
  {766} (\bibinfo {year} {2007})}\BibitemShut {NoStop}%
\bibitem [{\citenamefont {Knez}\ \emph {et~al.}(2011)\citenamefont {Knez},
  \citenamefont {Du},\ and\ \citenamefont {Sullivan}}]{knez2011}%
  \BibitemOpen
  \bibfield  {author} {\bibinfo {author} {\bibfnamefont {I.}~\bibnamefont
  {Knez}}, \bibinfo {author} {\bibfnamefont {R.-R.}\ \bibnamefont {Du}}, \ and\
  \bibinfo {author} {\bibfnamefont {G.}~\bibnamefont {Sullivan}},\ }\href@noop
  {} {\bibfield  {journal} {\bibinfo  {journal} {Physical review letters}\
  }\textbf {\bibinfo {volume} {107}},\ \bibinfo {pages} {136603} (\bibinfo
  {year} {2011})}\BibitemShut {NoStop}%
\bibitem [{\citenamefont {Weng}\ \emph
  {et~al.}(2014{\natexlab{a}})\citenamefont {Weng}, \citenamefont {Dai},\ and\
  \citenamefont {Fang}}]{MRS:9383312}%
  \BibitemOpen
  \bibfield  {author} {\bibinfo {author} {\bibfnamefont {H.}~\bibnamefont
  {Weng}}, \bibinfo {author} {\bibfnamefont {X.}~\bibnamefont {Dai}}, \ and\
  \bibinfo {author} {\bibfnamefont {Z.}~\bibnamefont {Fang}},\ }\href {\doibase
  10.1557/mrs.2014.216} {\bibfield  {journal} {\bibinfo  {journal} {MRS
  Bulletin}\ }\textbf {\bibinfo {volume} {39}},\ \bibinfo {pages} {849}
  (\bibinfo {year} {2014}{\natexlab{a}})}\BibitemShut {NoStop}%
\bibitem [{\citenamefont {Liu}\ \emph {et~al.}(2011)\citenamefont {Liu},
  \citenamefont {Feng},\ and\ \citenamefont {Yao}}]{liu2011}%
  \BibitemOpen
  \bibfield  {author} {\bibinfo {author} {\bibfnamefont {C.-C.}\ \bibnamefont
  {Liu}}, \bibinfo {author} {\bibfnamefont {W.}~\bibnamefont {Feng}}, \ and\
  \bibinfo {author} {\bibfnamefont {Y.}~\bibnamefont {Yao}},\ }\href@noop {}
  {\bibfield  {journal} {\bibinfo  {journal} {Physical review letters}\
  }\textbf {\bibinfo {volume} {107}},\ \bibinfo {pages} {076802} (\bibinfo
  {year} {2011})}\BibitemShut {NoStop}%
\bibitem [{\citenamefont {Xu}\ \emph {et~al.}(2013)\citenamefont {Xu},
  \citenamefont {Yan}, \citenamefont {Zhang}, \citenamefont {Wang},
  \citenamefont {Xu}, \citenamefont {Tang}, \citenamefont {Duan},\ and\
  \citenamefont {Zhang}}]{PhysRevLett.111.136804}%
  \BibitemOpen
  \bibfield  {author} {\bibinfo {author} {\bibfnamefont {Y.}~\bibnamefont
  {Xu}}, \bibinfo {author} {\bibfnamefont {B.}~\bibnamefont {Yan}}, \bibinfo
  {author} {\bibfnamefont {H.-J.}\ \bibnamefont {Zhang}}, \bibinfo {author}
  {\bibfnamefont {J.}~\bibnamefont {Wang}}, \bibinfo {author} {\bibfnamefont
  {G.}~\bibnamefont {Xu}}, \bibinfo {author} {\bibfnamefont {P.}~\bibnamefont
  {Tang}}, \bibinfo {author} {\bibfnamefont {W.}~\bibnamefont {Duan}}, \ and\
  \bibinfo {author} {\bibfnamefont {S.-C.}\ \bibnamefont {Zhang}},\ }\href
  {\doibase 10.1103/PhysRevLett.111.136804} {\bibfield  {journal} {\bibinfo
  {journal} {Phys. Rev. Lett.}\ }\textbf {\bibinfo {volume} {111}},\ \bibinfo
  {pages} {136804} (\bibinfo {year} {2013})}\BibitemShut {NoStop}%
\bibitem [{\citenamefont {Si}\ \emph {et~al.}(2014)\citenamefont {Si},
  \citenamefont {Liu}, \citenamefont {Xu}, \citenamefont {Wu}, \citenamefont
  {Gu},\ and\ \citenamefont {Duan}}]{PhysRevB.89.115429}%
  \BibitemOpen
  \bibfield  {author} {\bibinfo {author} {\bibfnamefont {C.}~\bibnamefont
  {Si}}, \bibinfo {author} {\bibfnamefont {J.}~\bibnamefont {Liu}}, \bibinfo
  {author} {\bibfnamefont {Y.}~\bibnamefont {Xu}}, \bibinfo {author}
  {\bibfnamefont {J.}~\bibnamefont {Wu}}, \bibinfo {author} {\bibfnamefont
  {B.-L.}\ \bibnamefont {Gu}}, \ and\ \bibinfo {author} {\bibfnamefont
  {W.}~\bibnamefont {Duan}},\ }\href {\doibase 10.1103/PhysRevB.89.115429}
  {\bibfield  {journal} {\bibinfo  {journal} {Phys. Rev. B}\ }\textbf {\bibinfo
  {volume} {89}},\ \bibinfo {pages} {115429} (\bibinfo {year}
  {2014})}\BibitemShut {NoStop}%
\bibitem [{\citenamefont {{Song}}\ \emph {et~al.}(2014)\citenamefont {{Song}},
  \citenamefont {{Liu}}, \citenamefont {{Yang}}, \citenamefont {{Han}},
  \citenamefont {{Ye}}, \citenamefont {{Fu}}, \citenamefont {{Yang}},
  \citenamefont {{Niu}}, \citenamefont {{Lu}},\ and\ \citenamefont
  {{Yao}}}]{2014arXiv1402.2399S}%
  \BibitemOpen
  \bibfield  {author} {\bibinfo {author} {\bibfnamefont {Z.}~\bibnamefont
  {{Song}}}, \bibinfo {author} {\bibfnamefont {C.-C.}\ \bibnamefont {{Liu}}},
  \bibinfo {author} {\bibfnamefont {J.}~\bibnamefont {{Yang}}}, \bibinfo
  {author} {\bibfnamefont {J.}~\bibnamefont {{Han}}}, \bibinfo {author}
  {\bibfnamefont {M.}~\bibnamefont {{Ye}}}, \bibinfo {author} {\bibfnamefont
  {B.}~\bibnamefont {{Fu}}}, \bibinfo {author} {\bibfnamefont {Y.}~\bibnamefont
  {{Yang}}}, \bibinfo {author} {\bibfnamefont {Q.}~\bibnamefont {{Niu}}},
  \bibinfo {author} {\bibfnamefont {J.}~\bibnamefont {{Lu}}}, \ and\ \bibinfo
  {author} {\bibfnamefont {Y.}~\bibnamefont {{Yao}}},\ }\href@noop {}
  {\bibfield  {journal} {\bibinfo  {journal} {{arXiv:1402.2399} [cond-mat]}\ }
  (\bibinfo {year} {2014})},\ \Eprint {http://arxiv.org/abs/1402.2399}
  {arXiv:1402.2399 [cond-mat.mtrl-sci]} \BibitemShut {NoStop}%
\bibitem [{\citenamefont {Zhou}\ \emph {et~al.}(2014)\citenamefont {Zhou},
  \citenamefont {Feng}, \citenamefont {Liu}, \citenamefont {Guan},\ and\
  \citenamefont {Yao}}]{Zhou2014}%
  \BibitemOpen
  \bibfield  {author} {\bibinfo {author} {\bibfnamefont {J.-J.}\ \bibnamefont
  {Zhou}}, \bibinfo {author} {\bibfnamefont {W.}~\bibnamefont {Feng}}, \bibinfo
  {author} {\bibfnamefont {C.-C.}\ \bibnamefont {Liu}}, \bibinfo {author}
  {\bibfnamefont {S.}~\bibnamefont {Guan}}, \ and\ \bibinfo {author}
  {\bibfnamefont {Y.}~\bibnamefont {Yao}},\ }\href {\doibase 10.1021/nl501907g}
  {\bibfield  {journal} {\bibinfo  {journal} {Nano Letters}\ }\textbf {\bibinfo
  {volume} {14}},\ \bibinfo {pages} {4767} (\bibinfo {year}
  {2014})}\BibitemShut {NoStop}%
\bibitem [{\citenamefont {Weng}\ \emph
  {et~al.}(2014{\natexlab{b}})\citenamefont {Weng}, \citenamefont {Dai},\ and\
  \citenamefont {Fang}}]{weng2014}%
  \BibitemOpen
  \bibfield  {author} {\bibinfo {author} {\bibfnamefont {H.}~\bibnamefont
  {Weng}}, \bibinfo {author} {\bibfnamefont {X.}~\bibnamefont {Dai}}, \ and\
  \bibinfo {author} {\bibfnamefont {Z.}~\bibnamefont {Fang}},\ }\href@noop {}
  {\bibfield  {journal} {\bibinfo  {journal} {Physical Review X}\ }\textbf
  {\bibinfo {volume} {4}},\ \bibinfo {pages} {011002} (\bibinfo {year}
  {2014}{\natexlab{b}})}\BibitemShut {NoStop}%
\bibitem [{\citenamefont {Luo}\ and\ \citenamefont {Xiang}(0)}]{XiangSQ}%
  \BibitemOpen
  \bibfield  {author} {\bibinfo {author} {\bibfnamefont {W.}~\bibnamefont
  {Luo}}\ and\ \bibinfo {author} {\bibfnamefont {H.}~\bibnamefont {Xiang}},\
  }\href {\doibase 10.1021/acs.nanolett.5b00418} {\bibfield  {journal}
  {\bibinfo  {journal} {Nano Letters}\ }\textbf {\bibinfo {volume} {0}},\
  \bibinfo {pages} {null} (\bibinfo {year} {0})},\ \Eprint
  {http://arxiv.org/abs/http://dx.doi.org/10.1021/acs.nanolett.5b00418}
  {http://dx.doi.org/10.1021/acs.nanolett.5b00418} \BibitemShut {NoStop}%
\bibitem [{\citenamefont {Bl{\"o}chl}(1994)}]{blochl1994}%
  \BibitemOpen
  \bibfield  {author} {\bibinfo {author} {\bibfnamefont {P.~E.}\ \bibnamefont
  {Bl{\"o}chl}},\ }\href@noop {} {\bibfield  {journal} {\bibinfo  {journal}
  {Physical Review B}\ }\textbf {\bibinfo {volume} {50}},\ \bibinfo {pages}
  {17953} (\bibinfo {year} {1994})}\BibitemShut {NoStop}%
\bibitem [{\citenamefont {Kresse}\ and\ \citenamefont
  {Joubert}(1999)}]{kresse1999}%
  \BibitemOpen
  \bibfield  {author} {\bibinfo {author} {\bibfnamefont {G.}~\bibnamefont
  {Kresse}}\ and\ \bibinfo {author} {\bibfnamefont {D.}~\bibnamefont
  {Joubert}},\ }\href@noop {} {\bibfield  {journal} {\bibinfo  {journal}
  {Physical Review B}\ }\textbf {\bibinfo {volume} {59}},\ \bibinfo {pages}
  {1758} (\bibinfo {year} {1999})}\BibitemShut {NoStop}%
\bibitem [{\citenamefont {Kresse}\ and\ \citenamefont
  {Furthm{\"u}ller}(1996{\natexlab{a}})}]{kresse1996_1}%
  \BibitemOpen
  \bibfield  {author} {\bibinfo {author} {\bibfnamefont {G.}~\bibnamefont
  {Kresse}}\ and\ \bibinfo {author} {\bibfnamefont {J.}~\bibnamefont
  {Furthm{\"u}ller}},\ }\href@noop {} {\bibfield  {journal} {\bibinfo
  {journal} {Computational Materials Science}\ }\textbf {\bibinfo {volume}
  {6}},\ \bibinfo {pages} {15} (\bibinfo {year}
  {1996}{\natexlab{a}})}\BibitemShut {NoStop}%
\bibitem [{\citenamefont {Kresse}\ and\ \citenamefont
  {Furthm{\"u}ller}(1996{\natexlab{b}})}]{kresse1996_2}%
  \BibitemOpen
  \bibfield  {author} {\bibinfo {author} {\bibfnamefont {G.}~\bibnamefont
  {Kresse}}\ and\ \bibinfo {author} {\bibfnamefont {J.}~\bibnamefont
  {Furthm{\"u}ller}},\ }\href@noop {} {\bibfield  {journal} {\bibinfo
  {journal} {Physical Review B}\ }\textbf {\bibinfo {volume} {54}},\ \bibinfo
  {pages} {11169} (\bibinfo {year} {1996}{\natexlab{b}})}\BibitemShut {NoStop}%
\bibitem [{\citenamefont {Perdew}\ \emph {et~al.}(1996)\citenamefont {Perdew},
  \citenamefont {Burke},\ and\ \citenamefont {Ernzerhof}}]{Perdew1996}%
  \BibitemOpen
  \bibfield  {author} {\bibinfo {author} {\bibfnamefont {J.}~\bibnamefont
  {Perdew}}, \bibinfo {author} {\bibfnamefont {K.}~\bibnamefont {Burke}}, \
  and\ \bibinfo {author} {\bibfnamefont {M.}~\bibnamefont {Ernzerhof}},\
  }\href@noop {} {\bibfield  {journal} {\bibinfo  {journal} {Phys. Rev. Lett.}\
  }\textbf {\bibinfo {volume} {77}},\ \bibinfo {pages} {3865} (\bibinfo {year}
  {1996})}\BibitemShut {NoStop}%
\bibitem [{\citenamefont {Togo}\ \emph {et~al.}(2008)\citenamefont {Togo},
  \citenamefont {Oba},\ and\ \citenamefont {Tanaka}}]{togo2008}%
  \BibitemOpen
  \bibfield  {author} {\bibinfo {author} {\bibfnamefont {A.}~\bibnamefont
  {Togo}}, \bibinfo {author} {\bibfnamefont {F.}~\bibnamefont {Oba}}, \ and\
  \bibinfo {author} {\bibfnamefont {I.}~\bibnamefont {Tanaka}},\ }\href@noop {}
  {\bibfield  {journal} {\bibinfo  {journal} {Physical Review B}\ }\textbf
  {\bibinfo {volume} {78}},\ \bibinfo {pages} {134106} (\bibinfo {year}
  {2008})}\BibitemShut {NoStop}%
\bibitem [{\citenamefont {Marzari}\ and\ \citenamefont
  {Vanderbilt}(1997)}]{marzari1997}%
  \BibitemOpen
  \bibfield  {author} {\bibinfo {author} {\bibfnamefont {N.}~\bibnamefont
  {Marzari}}\ and\ \bibinfo {author} {\bibfnamefont {D.}~\bibnamefont
  {Vanderbilt}},\ }\href@noop {} {\bibfield  {journal} {\bibinfo  {journal}
  {Physical review B}\ }\textbf {\bibinfo {volume} {56}},\ \bibinfo {pages}
  {12847} (\bibinfo {year} {1997})}\BibitemShut {NoStop}%
\bibitem [{\citenamefont {Souza}\ \emph {et~al.}(2001)\citenamefont {Souza},
  \citenamefont {Marzari},\ and\ \citenamefont {Vanderbilt}}]{souza2001}%
  \BibitemOpen
  \bibfield  {author} {\bibinfo {author} {\bibfnamefont {I.}~\bibnamefont
  {Souza}}, \bibinfo {author} {\bibfnamefont {N.}~\bibnamefont {Marzari}}, \
  and\ \bibinfo {author} {\bibfnamefont {D.}~\bibnamefont {Vanderbilt}},\
  }\href@noop {} {\bibfield  {journal} {\bibinfo  {journal} {Physical Review
  B}\ }\textbf {\bibinfo {volume} {65}},\ \bibinfo {pages} {035109} (\bibinfo
  {year} {2001})}\BibitemShut {NoStop}%
\bibitem [{\citenamefont {Mostofi}\ \emph {et~al.}(2008)\citenamefont
  {Mostofi}, \citenamefont {Yates}, \citenamefont {Lee}, \citenamefont {Souza},
  \citenamefont {Vanderbilt},\ and\ \citenamefont {Marzari}}]{mostofi2008}%
  \BibitemOpen
  \bibfield  {author} {\bibinfo {author} {\bibfnamefont {A.~A.}\ \bibnamefont
  {Mostofi}}, \bibinfo {author} {\bibfnamefont {J.~R.}\ \bibnamefont {Yates}},
  \bibinfo {author} {\bibfnamefont {Y.-S.}\ \bibnamefont {Lee}}, \bibinfo
  {author} {\bibfnamefont {I.}~\bibnamefont {Souza}}, \bibinfo {author}
  {\bibfnamefont {D.}~\bibnamefont {Vanderbilt}}, \ and\ \bibinfo {author}
  {\bibfnamefont {N.}~\bibnamefont {Marzari}},\ }\href@noop {} {\bibfield
  {journal} {\bibinfo  {journal} {Computer physics communications}\ }\textbf
  {\bibinfo {volume} {178}},\ \bibinfo {pages} {685} (\bibinfo {year}
  {2008})}\BibitemShut {NoStop}%
\bibitem [{\citenamefont {Sancho}\ \emph {et~al.}(1984)\citenamefont {Sancho},
  \citenamefont {Sancho},\ and\ \citenamefont {Rubio}}]{sancho1984quick}%
  \BibitemOpen
  \bibfield  {author} {\bibinfo {author} {\bibfnamefont {M.~L.}\ \bibnamefont
  {Sancho}}, \bibinfo {author} {\bibfnamefont {J.~L.}\ \bibnamefont {Sancho}},
  \ and\ \bibinfo {author} {\bibfnamefont {J.}~\bibnamefont {Rubio}},\
  }\href@noop {} {\bibfield  {journal} {\bibinfo  {journal} {Journal of Physics
  F: Metal Physics}\ }\textbf {\bibinfo {volume} {14}},\ \bibinfo {pages}
  {1205} (\bibinfo {year} {1984})}\BibitemShut {NoStop}%
\bibitem [{\citenamefont {Sancho}\ \emph {et~al.}(1985)\citenamefont {Sancho},
  \citenamefont {Sancho}, \citenamefont {Sancho},\ and\ \citenamefont
  {Rubio}}]{sancho1985highly}%
  \BibitemOpen
  \bibfield  {author} {\bibinfo {author} {\bibfnamefont {M.~L.}\ \bibnamefont
  {Sancho}}, \bibinfo {author} {\bibfnamefont {J.~L.}\ \bibnamefont {Sancho}},
  \bibinfo {author} {\bibfnamefont {J.~L.}\ \bibnamefont {Sancho}}, \ and\
  \bibinfo {author} {\bibfnamefont {J.}~\bibnamefont {Rubio}},\ }\href@noop {}
  {\bibfield  {journal} {\bibinfo  {journal} {Journal of Physics F: Metal
  Physics}\ }\textbf {\bibinfo {volume} {15}},\ \bibinfo {pages} {851}
  (\bibinfo {year} {1985})}\BibitemShut {NoStop}%
\bibitem [{\citenamefont {Fu}\ and\ \citenamefont {Kane}(2007)}]{Fu:2007ei}%
  \BibitemOpen
  \bibfield  {author} {\bibinfo {author} {\bibfnamefont {L.}~\bibnamefont
  {Fu}}\ and\ \bibinfo {author} {\bibfnamefont {C.}~\bibnamefont {Kane}},\
  }\href@noop {} {\bibfield  {journal} {\bibinfo  {journal} {Physical Review
  B}\ }\textbf {\bibinfo {volume} {76}},\ \bibinfo {pages} {045302} (\bibinfo
  {year} {2007})}\BibitemShut {NoStop}%
\bibitem [{\citenamefont {Heyd}\ \emph {et~al.}(2003)\citenamefont {Heyd},
  \citenamefont {Scuseria},\ and\ \citenamefont {Ernzerhof}}]{heyd2003hybrid}%
  \BibitemOpen
  \bibfield  {author} {\bibinfo {author} {\bibfnamefont {J.}~\bibnamefont
  {Heyd}}, \bibinfo {author} {\bibfnamefont {G.~E.}\ \bibnamefont {Scuseria}},
  \ and\ \bibinfo {author} {\bibfnamefont {M.}~\bibnamefont {Ernzerhof}},\
  }\href@noop {} {\bibfield  {journal} {\bibinfo  {journal} {The Journal of
  Chemical Physics}\ }\textbf {\bibinfo {volume} {118}},\ \bibinfo {pages}
  {8207} (\bibinfo {year} {2003})}\BibitemShut {NoStop}%
\bibitem [{\citenamefont {Heyd}\ \emph {et~al.}(2006)\citenamefont {Heyd},
  \citenamefont {Scuseria},\ and\ \citenamefont {Ernzerhof}}]{heyd2006hybrid}%
  \BibitemOpen
  \bibfield  {author} {\bibinfo {author} {\bibfnamefont {J.}~\bibnamefont
  {Heyd}}, \bibinfo {author} {\bibfnamefont {G.~E.}\ \bibnamefont {Scuseria}},
  \ and\ \bibinfo {author} {\bibfnamefont {M.}~\bibnamefont {Ernzerhof}},\
  }\href@noop {} {\bibfield  {journal} {\bibinfo  {journal} {The Journal of
  Chemical Physics}\ }\textbf {\bibinfo {volume} {124}},\ \bibinfo {pages}
  {219906} (\bibinfo {year} {2006})}\BibitemShut {NoStop}%
\end{thebibliography}%

\newpage
\section{Appendix: Other TMD Haeckelites MX$_2$-4-8}
\begin{figure}[htp]
\includegraphics[width=5.2in]{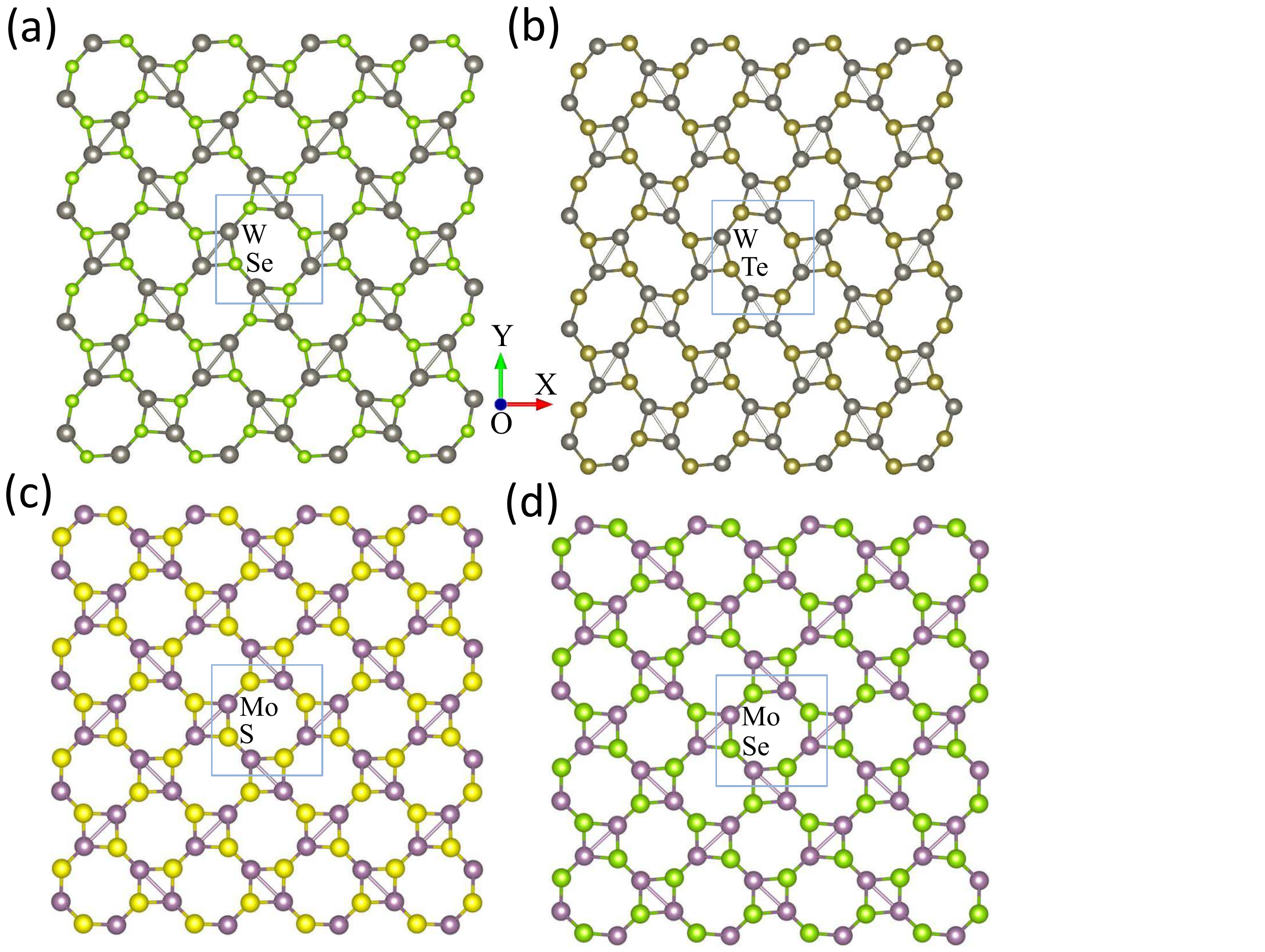}
      \caption{(color online). Top views of crystal structure of WSe$_2$-4-8 (a), WTe$_2$-4-8 (a), MoS$_2$-4-8 (c) and MoSe$_2$-4-8 (d).}
\end{figure}

\begin{figure}[htp]
\includegraphics[width=5.2in]{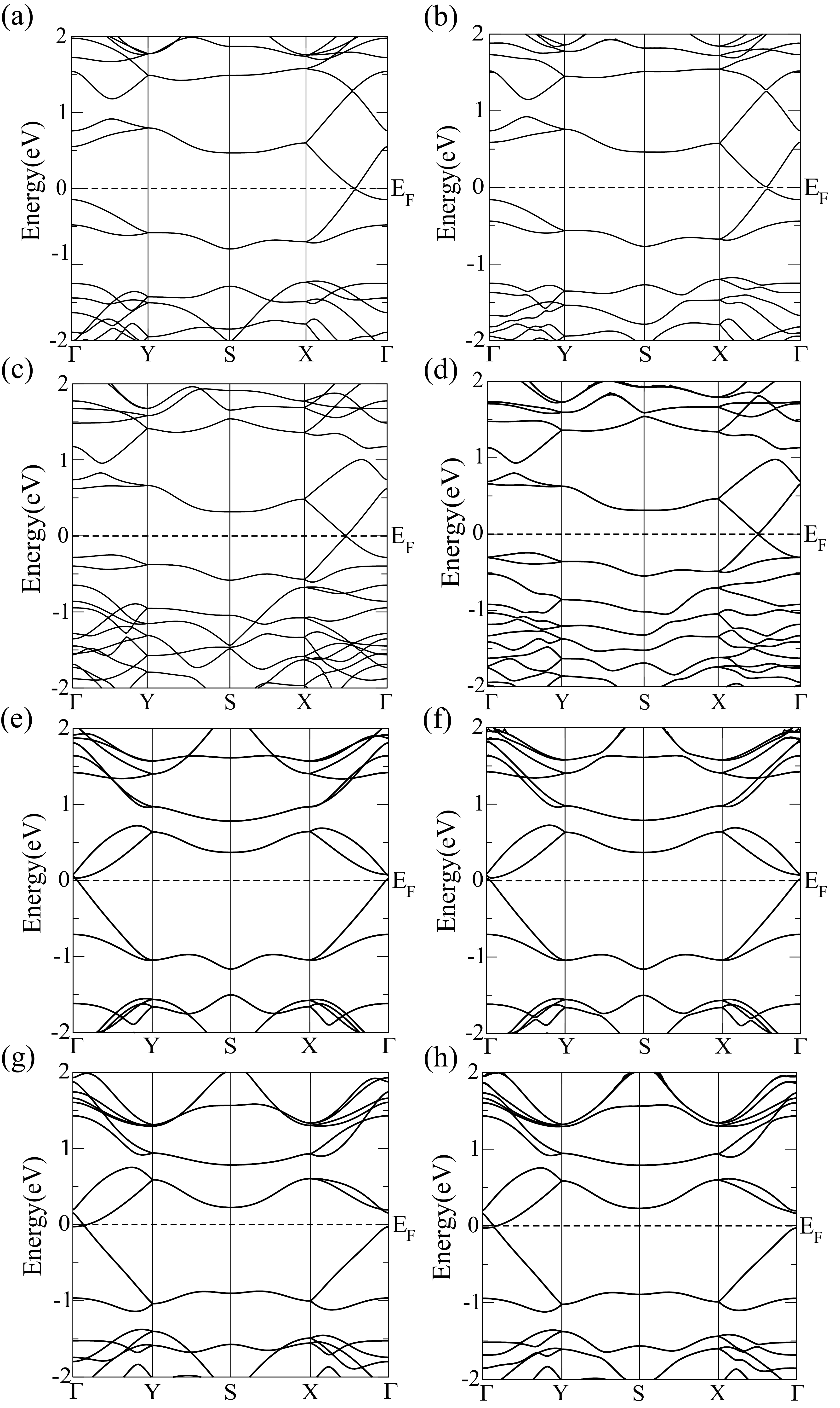}
      \caption{(color online). Band structures of WSe$_2$-4-8 (a, b), WTe$_2$-4-8 (c, d), MoS$_2$-4-8 (e, f) and MoSe$_2$-4-8
      (g, h). Left panel for non-SOC case and right panel for SOC case.}
\end{figure}

\begin{figure}[htp]
\includegraphics[width=5.2in]{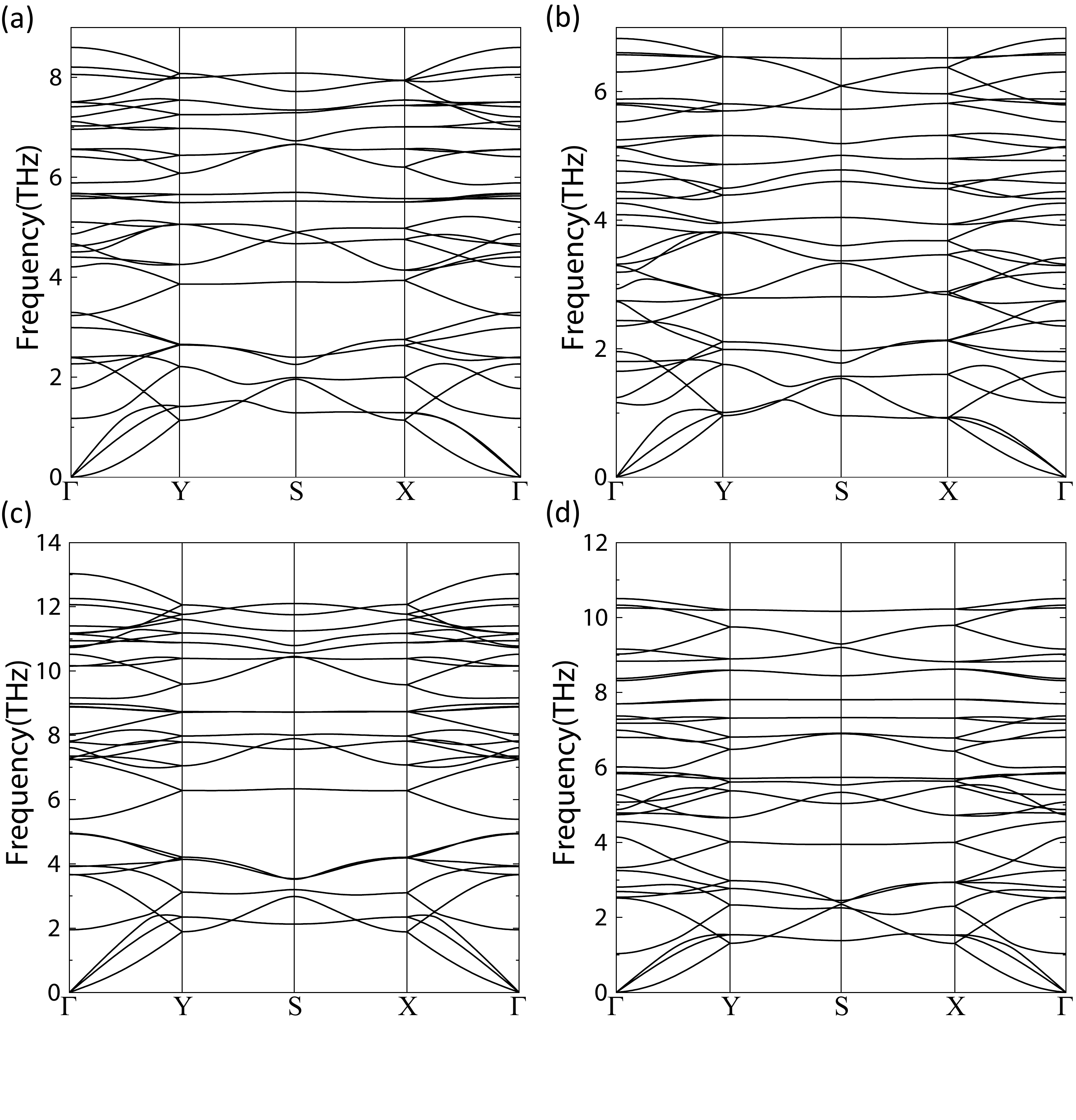}
      \caption{(color online). Phonon dispersions of WSe$_2$-4-8 (a), WTe$_2$-4-8 (b), MoS$_2$-4-8 (c) and MoSe$_2$-4-8 (d).}
\end{figure}

\begin{figure}[htp]
\includegraphics[width=5.1in]{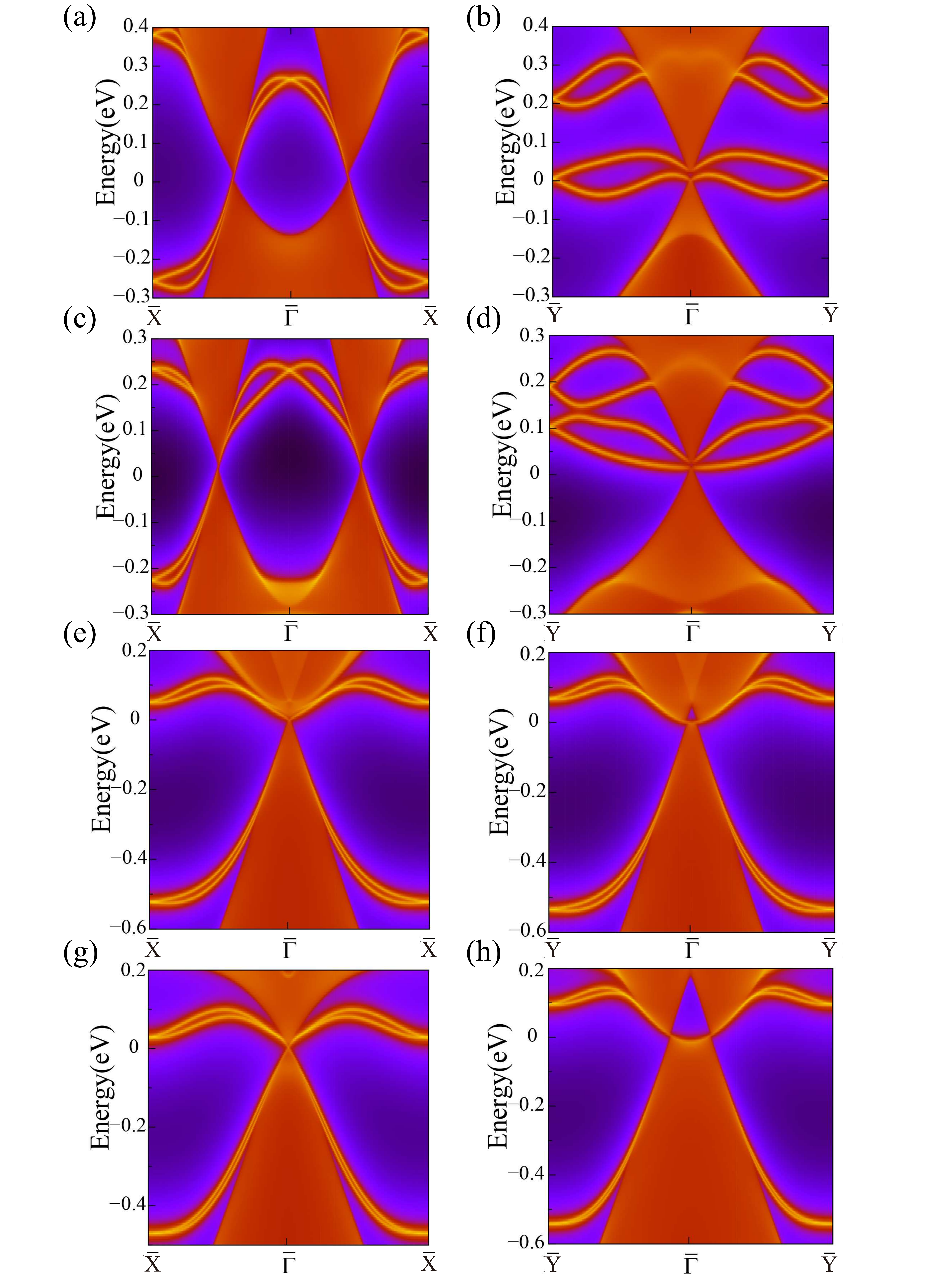}
      \caption{(color online). Edge states for WSe$_2$-4-8 (a, b), WTe$_2$-4-8 (c, d), MoS$_2$-4-8 (e, f) and MoSe$_2$-4-8 (g, h).
       Left panel for X edge and right panel for Y edge.}
\end{figure}

\end{document}